\documentclass[preprint,12pt]{elsarticle}

\usepackage{epsfig}

\journal{Nucl. Instrum. Meth. Phys. Res. A}

\begin{document}

\begin{frontmatter}

\title{First scintillating bolometer tests of a CLYMENE R\&D on Li$_2$MoO$_4$ scintillators towards a large-scale double-beta decay experiment}

\author[ICMCB]{G.~Bu\c{s}e}
\author[CSNSM,DISAT]{A.~Giuliani}
\author[CSNSM]{P.~de~Marcillac}
\author[CSNSM]{S.~Marnieros}
\author[CEA-IRFU]{C.~Nones}
\author[CSNSM]{V.~Novati}
\author[CSNSM]{E.~Olivieri}
\author[CSNSM,KINR]{D.V.~Poda\corref{e1}}
 \cortext[e1]{Corresponding author}
 \ead{denys.poda@csnsm.in2p3.fr}
\author[CSNSM]{T.~Redon}
\author[ICMCB]{J.-B.~Sand}
\author[ICMCB,UCBL]{P.~Veber}
\author[ICMCB,UGA]{M.~Vel\'azquez}
\author[CEA-IRFU]{A.S.~Zolotarova}

\address[ICMCB]{ICMCB, UMR 5026, CNRS-Universit\'{e} de Bordeaux-INP, 33608 Pessac Cedex, France}
\address[CSNSM]{CSNSM, Univ. Paris-Sud, CNRS/IN2P3, Universit\'e Paris-Saclay, 91405 Orsay, France}
\address[DISAT]{DISAT, Universit\`a dell'Insubria, 22100 Como, Italy}
\address[CEA-IRFU]{IRFU, CEA, Universit\'{e} Paris-Saclay, F-91191 Gif-sur-Yvette, France}
\address[KINR]{Institute for Nuclear Research, 03028 Kyiv, Ukraine}
\address[UGA]{Universit\'e Grenoble Alpes, CNRS, Grenoble INP, SIMAP, 38402 Saint Martin d'H\'eres, France}
\address[UCBL]{Universit\'e Lyon, Universit\'e Claude Bernard Lyon 1, CNRS, Institut Lumi\'ere Mati\'ere, UMR 5306, 69100 Villeurbanne, France}

\begin{abstract}
A new R\&D on lithium molybdate scintillators has begun within a project CLYMENE (Czochralski growth of Li$_2$MoO$_4$ crYstals for the scintillating boloMeters used in the rare EveNts sEarches). One of the main goals of the CLYMENE is a realization of a Li$_2$MoO$_4$ crystal growth line to be complementary to the one recently developed by LUMINEU in view of a mass production capacity for CUPID, a next-generation tonne-scale bolometric experiment to search for neutrinoless double-beta decay. In the present paper we report the investigation of performance and radiopurity of 158-g and 13.5-g scintillating bolometers based on a first large-mass (230~g) Li$_2$MoO$_4$ crystal scintillator developed within the CLYMENE project. In particular, a good energy resolution (2--7 keV FWHM in the energy range of 0.2--5~MeV), one of the highest light yield (0.97~keV/MeV) amongst Li$_2$MoO$_4$ scintillating bolometers, an efficient alpha particles discrimination (10$\sigma$) and potentially low internal radioactive contamination (below 0.2--0.3~mBq/kg of U/Th, but 1.4~mBq/kg of $^{210}$Po) demonstrate prospects of the CLYMENE in the development of high quality and radiopure Li$_2$MoO$_4$ scintillators for CUPID.   
\end{abstract}


\end{frontmatter}


\section{Introduction}
\label{sec:intro}

A unique probe of the new physics beyond the Standard Model will be realized in the upcoming decade by several next-generation experiments, in particular CUPID (CUORE Upgrade with Particle IDentification) \cite{Wang:2015a}, aiming at the detection of neutrinoless double-beta decay \cite{Vergados:2016}. The basic option for CUPID is to exploit the infrastructure of the recently started CUORE experiment with a tonne-scale bolometer array \cite{Alduino:2017}. Extensive R\&D activities \cite{Wang:2015b,Poda:2017a} are presently ongoing to develop a bolometric technique which can accomplish the CUPID goals \cite{Wang:2015a}. The key points addressed to a bolometric technique to be applied in CUPID are reproducible crystallization and detector technologies satisfying the project requirements on the production, purity, and performance. 

A technology of a mass production of high quality radiopure natural and $^{100}$Mo-enriched lithium molybdate scintillators and their use as high performance scintillating bolometers has been recently developed \cite{Armengaud:2017} within a project LUMINEU (Luminescent Underground Molybdenum Investigation for NEUtrino mass and nature), considered as a part of CUPID R\&D. In view of the necessity of the material mass production at the tonne-scale level in a strict time period, it is crucial to diversify the capacity of crystal growth for CUPID. For that reason, a project CLYMENE (Czochralski growth of Li$_2$MoO$_4$ crYstals for the scintillating boloMeters used in the rare EveNts sEarches), recently funded by ANR (France), is well suitable to provide $\sim$0.5-kg Li$_2$MoO$_4$ crystals production line complementary to the existing one of the LUMINEU \cite{Armengaud:2017}. This ambitious goal is going to be achieved within CLYMENE by means of both combined Czochralski pulling and modeling, Li$_2$MoO$_4$ single crystals characterizations and exploratory bolometer tests according to well-established LUMINEU detector protocol. The present paper is devoted to the bolometric investigation of a large and a small scintillation elements produced from 230-g crystal boule, a first large-volume scintillator developed within the CLYMENE program \cite{Velazquez:2017}. The encouraging results detailed below unambiguously demonstrate a good starting point for the successful realization of the CLYMENE project.

\section{Bolometric tests of Li$_2$MoO$_4$ crystals}
\label{sec:test}

In the present study we characterized two colorless transparent scintillation elements produced from the 0.23~kg Li$_2$MoO$_4$ crystal scintillator, grown by the Czochralski method at ICMCB (Bordeaux, France) from 5N5 MoO$_3$ and 5N Li$_2$CO$_3$ powders \cite{Velazquez:2017}. A first sample, LMO-large ($\oslash$40$\times$40~mm, 158~g), had a cleavage as a result of an accident during the cut of the crystal boule. Thereby, it was naturally assumed that observed peculiarities of the signal shape together with poor bolometric properties of the LMO-large detector, briefly reported in \cite{Velazquez:2017} and detailed in section~\ref{sec:results}, were caused by the crystal internal crack. To confirm that, we repeated the measurements with another sample, LMO-small (28$\times$27$\times$6~mm, 13.5~g), cut from the part of the LMO-large element corresponding to the top part of the crystal boule, about 3~cm below the seed. The LMO-large crystal was cut parallel to the crack with a wire saw in order to identify what this crack exactly is. The single crystal cut parallel to the crack was oriented by the Laue method on a Delta Technologies International GM WS Series X-ray goniometer head. The Laue photographs were monitored and indexed by means of the Orient Express software. The orientation of the crack was found to be $\approx$11° $\Phi_Y$ and $\approx$1.1° $\Phi_Z$ from (001) planes. Moreover, the growth direction of the crystal was determined to be [110], which corresponds to the orientation of the seed \cite{Velazquez:2017}. An analysis of the chemical bonding in the (110) and (001) crystallographic planes clearly shows that the latter exhibit a much lower density of chemical bonds. This is consistent with the lattice spacings in the crystal structure, the largest one being d$_{[001]}$ = c $\approx$ 9.6~$\textrm{\AA}$ (c is a lattice parameter of the R-3 space group of the studied Li$_2$MoO$_4$ crystal, see details in \cite{Velazquez:2017}), which can be viewed as a weakness at the interatomic potential level.

The low temperature tests (at 13.5--18~mK) of both Li$_2$MoO$_4$ samples have been performed in an aboveground cryogenic laboratory at CSNSM (Orsay, France) with the help of a pulse-tube cryostat described in Ref. \cite{Mancuso:2014}. The LMO-large and LMO-small crystals were placed in the detector holders used for the bolometric test of the same-size Li$_2$MoO$_4$ \cite{Bekker:2016} and $^{116}$Cd-enriched CdWO$_4$ \cite{Barabash:2016} scintillating bolometers respectively; both also tested in the same set-up. Specifically, the samples were mounted inside a copper housing with the help of PTFE supporting elements and coupled to a high-purity grade Ge wafer ($\oslash$44$\times$0.175~mm) acting as a photodetector. The inner part of the detector holders was covered by a reflecting foil (Vikuiti\textsuperscript{\texttrademark} Enhanced Specular Reflector Film) to improve the collection of the emitted scintillation light. Both Li$_2$MoO$_4$ and Ge absorbers were equipped with an NTD (Neutron Transmutation Doped) Ge thermistor in order to detect particle interaction as a temperature rise. In spite of the same detection mechanism, we call Li$_2$MoO$_4$ and Ge bolometers as the heat and the light detectors respectively to distinguish their main purpose. In addition, a small Si:P chip \cite{Andreotti:2012} was epoxy-glued on the Li$_2$MoO$_4$ crystals to inject a constant power periodically for the off-line stabilization of the thermal gain, as it is detailed in \cite{Arnaboldi:2011b}. The detectors read-out was performed with a low-noise room-temperature DC electronics located in a Faraday cage. The stream data were acquired by 16-bit ADC with 10~kHz sampling rate. The heat detectors were calibrated by $\gamma$ quanta of natural radioactivity (mainly by $^{214}$Bi and $^{214}$Pb, daughters of $^{226}$Ra) and a low-activity $^{232}$Th source (LMO-large only). Both light detectors were permanently irradiated by X-rays of a $^{55}$Fe source (5.9 and 6.5~keV doublet of $^{55}$Mn). The data analysis was realized with the help of the optimum filter technique \cite{Gatti:1986}.

\section{Results and discussion}
\label{sec:results}

\begin{figure*}[htb]
\begin{center}
\resizebox{0.49\textwidth}{!}{\includegraphics{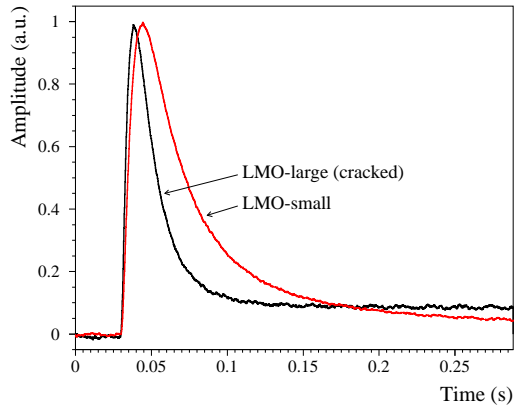}}
\end{center}
\caption{(Color online) The average pulses of the LMO-large (158~g) and LMO-small (13.5~g) bolometers constructed from the data accumulated in the same cryogenic set-up. The pulses are normalized on their corresponding maximum. A fast pulse-shape profile and a clear long-time component of the decaying part of the LMO-large signal are caused by the crystal crack (see text).}
 \label{fig:meanpulses}
\end{figure*}

In the bolometric tests with the LMO-large crystal we observed that the heat signals are untypically fast for the NTD-instrumented massive bolometer; e.g. 5~ms and 20~ms rise and decay time parameters are about factor of 2--6 and 4--20 respectively faster than those of similar size Li$_2$MoO$_4$-based cryogenic detectors \cite{Bekker:2016,Armengaud:2017}. Moreover, the decaying edge of the signals contains a prominent second component exhibited as a long plateau-like tail at the level of $\sim$10\% of the amplitude maximum (see Fig.~\ref{fig:meanpulses}). We assumed that such behavior of the LMO-large pulse-shape can be explained by the crystal crack which affects the phonons collection and the further thermalization to the heat bath temperature after the particle interaction. In the subsequent low temperature measurements with the LMO-small sample, cut from the perfect part of the LMO-large element, expected pulse-shape time constants and no indication of the plateau-like tail were found (see Fig.~\ref{fig:meanpulses}) confirming our assumption. 

\begin{table}[htb]
 \caption{Best performance of scintillating bolometers based on Ge light detectors (LDs) coupled to LMO-large and LMO-small heat detectors (HDs). Table contains the information about the detector sensitivity characterized as a voltage signal amplitude per unit of the deposited energy, the energy resolution (FWHM) of the baseline at the filter output, at 5.9 keV X-ray of $^{55}$Mn (LDs only), 609 keV $\gamma$ quanta of $^{214}$Bi and $\alpha$+triton peak induced by thermal neutron capture on $^{6}$Li (HDs only). Particle identification parameters as light yield for $\gamma$($\beta$)'s ($LY_{\gamma(\beta)}$) and quenching factor of light yield for $\alpha$ particles ($QF_{\alpha+t}$, estimated for $\alpha$+triton events), as well as  discrimination power ($DP_{\alpha+t/\gamma(\beta)}$) between $\alpha$+triton and $\gamma$($\beta$) distributions are also quoted. Some results reported in \cite{Velazquez:2017} for the LMO-large detector have been revised.}
\begin{center}
\begin{tabular}{lll}
 \hline
HD 		& LMO-large (158 g)	& LMO-small (13.5 g) \\
LD		& Ge ($\oslash$44-mm)	& Ge ($\oslash$44-mm) \\
 \hline
HD Signal (nV/keV)					& 2.9			& 202 \\
HD FWHM (keV) at baseline		& 32(1)		& 2.0(1) \\
HD FWHM (keV) at 609 keV		& 31(4)		& 1.8(6) \\
HD FWHM (keV) at 4784 keV		& 113(3)	& 6.7(5) \\
 \hline
LD Signal (nV/keV)					& 844			& 803 \\
LD FWHM (keV) at baseline		& 0.21(1)	& 0.29(1) \\
LD FWHM (keV) at 5.9 keV 		& 0.40(1)	& 0.32(1) \\
 \hline
$LY_{\gamma(\beta)}$ (keV/MeV)	& 0.973(1)	& 0.914(2) \\
$QF_{\alpha+t}$									& 0.23			& 0.24 \\
$DP_{\alpha+t/\gamma(\beta)}$		& 10.3			& 9.1 \\
 \hline
 \end{tabular}
  \label{tab:performance}
 \end{center}
 \end{table}

\begin{figure}[htb]
\begin{center}
\resizebox{0.49\textwidth}{!}{\includegraphics{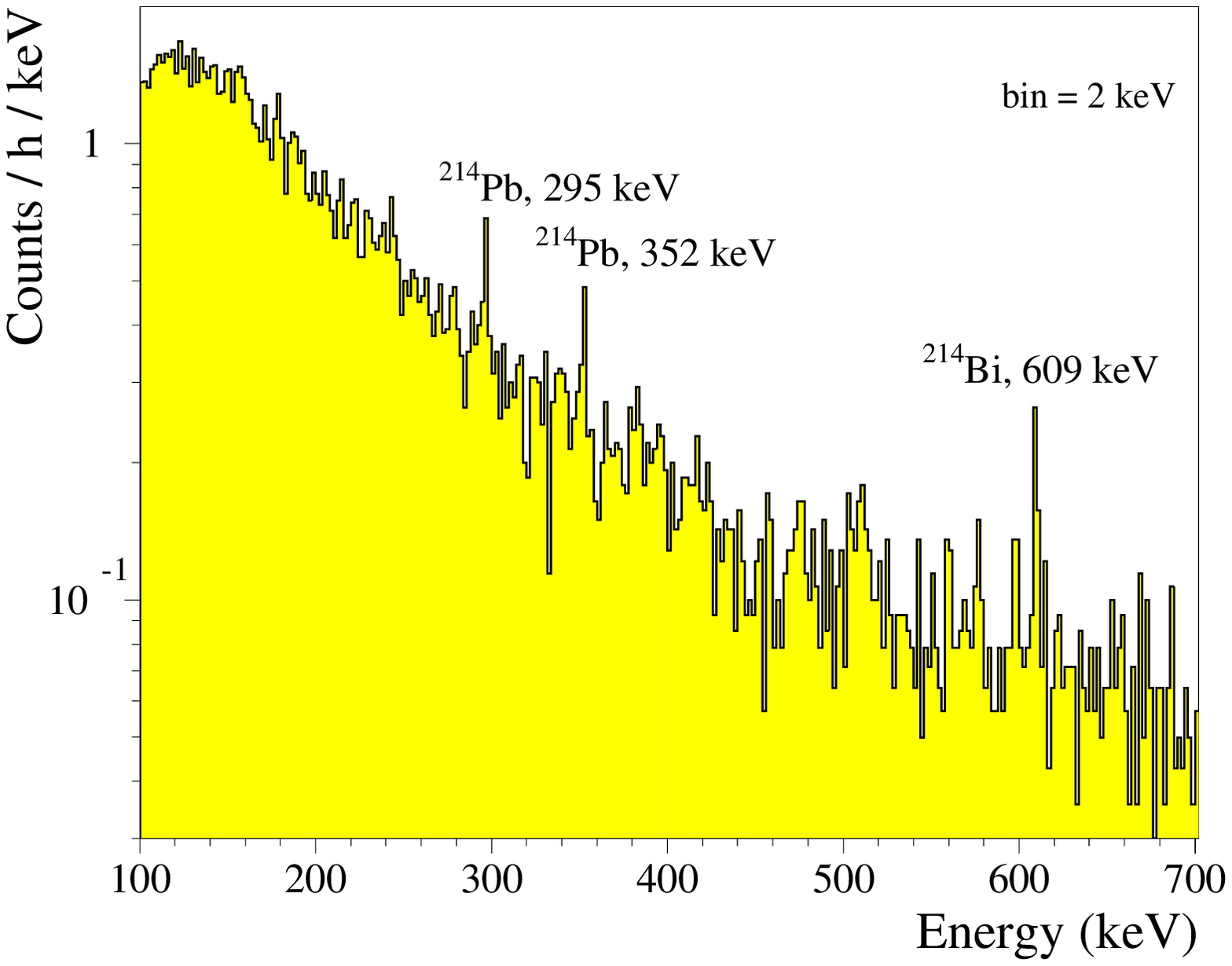}}
\resizebox{0.49\textwidth}{!}{\includegraphics{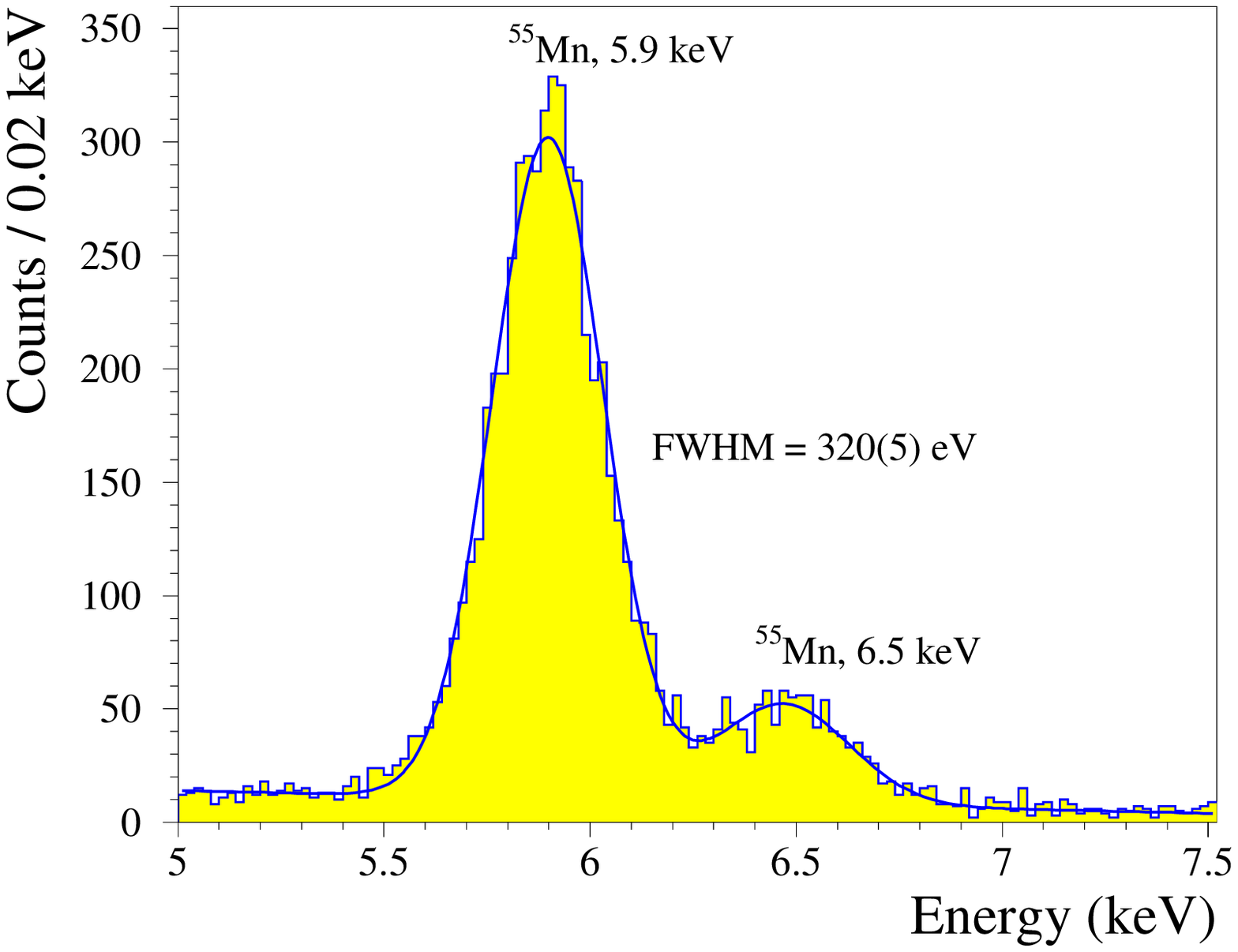}}
\end{center}
\caption{(Left) A fragment of the background energy spectrum accumulated by the LMO-small (13.5 g) cryogenic detector operated above ground over 70 h. (Right) The energy spectrum of the $^{55}$Fe X-ray source measured over 97.4~h by the Ge thin light detector ($\oslash$44$\times$0.175~mm) coupled to the LMO-small scintillating bolometer. The 5.9 and 6.5~keV doublet of $^{55}$Mn K$_\alpha$ and K$_\beta$ is almost resolved; the energy resolution is FWHM = (320$\pm$5)~eV. The fit to the spectrum is shown by a solid line.}
 \label{fig:calibr_spectra}
\end{figure}

Consequently of the LMO-large cleavage, this detector had also the affected performance namely a low signal sensitivity ($\sim$3~nV/keV) and a poor energy resolution (e.g. about 30 keV FWHM baseline noise), see Table~\ref{tab:performance}. Whereas, a large sensitivity ($\sim$200~nV/keV, a typical value for a small sample with a reduced heat capacity) and a high energy resolution (2 keV FWHM noise, typical for macrobolometers in the used cryogenic set-up) exhibit excellent performance of the LMO-small bolometer, not affected by the internal defect. In particular, the LMO-small performance is similar to that of the same-size $^{116}$CdWO$_4$ bolometer (135~nV/keV signal, 1.5~keV FWHM noise, 3.7~keV FWHM at 609 keV) \cite{Barabash:2016}. The energy spectrum of the background data of the LMO-small bolometer with a few gamma peaks originated to the environmental radioactivity is illustrated in Fig.~\ref{fig:calibr_spectra} (Left).

The best performances of Ge bolometers, used for the registration of the Li$_2$MoO$_4$ crystals scintillation, were rather similar demonstrating around 0.8~$\mu$V/keV signal and 0.2--0.3~keV FWHM baseline noise (see Table~\ref{tab:performance}); both parameters are typical for NTD-instrumented Ge light detectors operated in that pulse-tube cryostat. The energy spectrum of the $^{55}$Fe X-ray calibration source accumulated by the Ge light detector coupled to the LMO-small bolometer is shown in Fig.~\ref{fig:calibr_spectra} (Right).

\begin{figure*}
\begin{center}
\resizebox{0.49\textwidth}{!}{\includegraphics{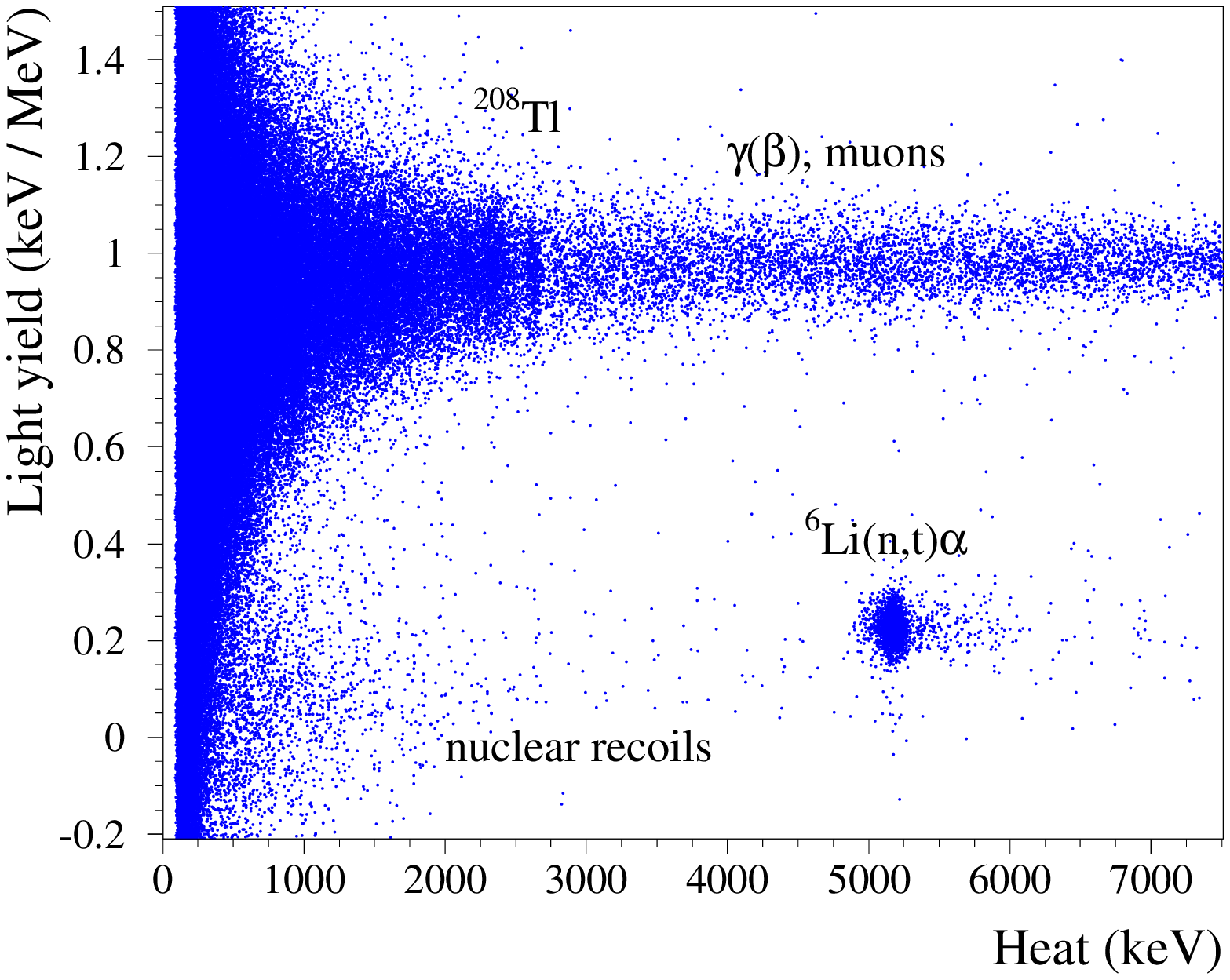}}
\resizebox{0.49\textwidth}{!}{\includegraphics{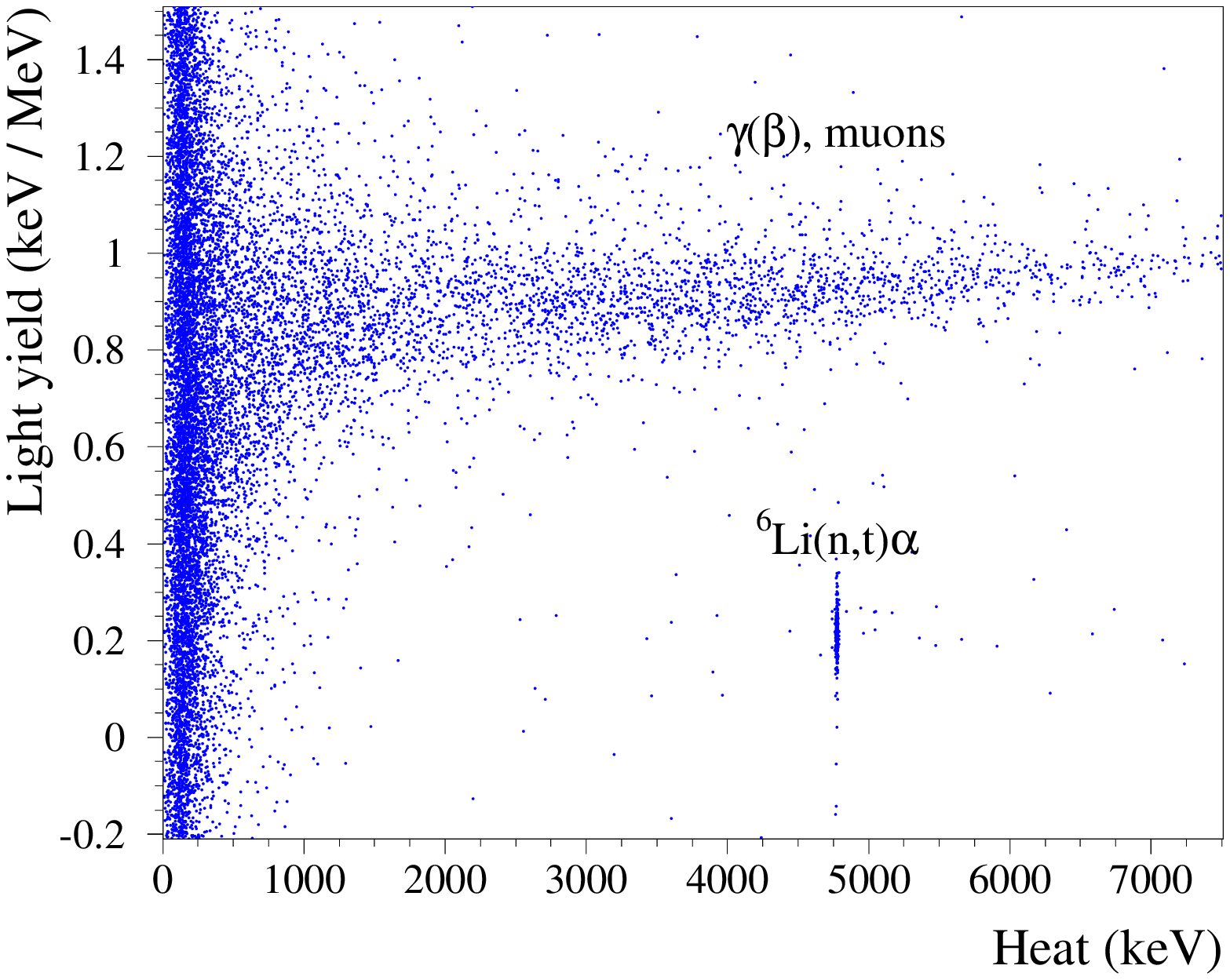}}
\resizebox{0.49\textwidth}{!}{\includegraphics{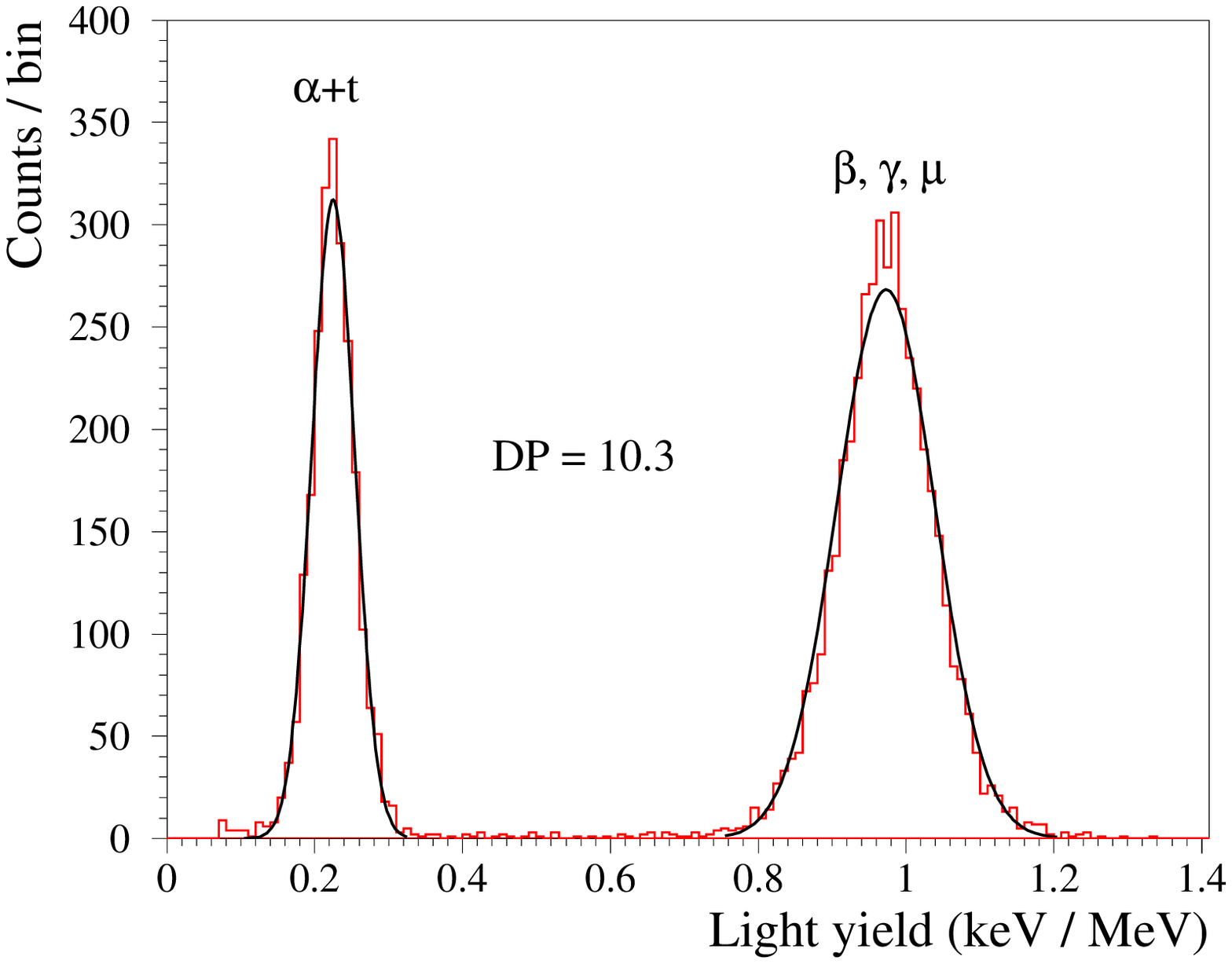}}
\resizebox{0.49\textwidth}{!}{\includegraphics{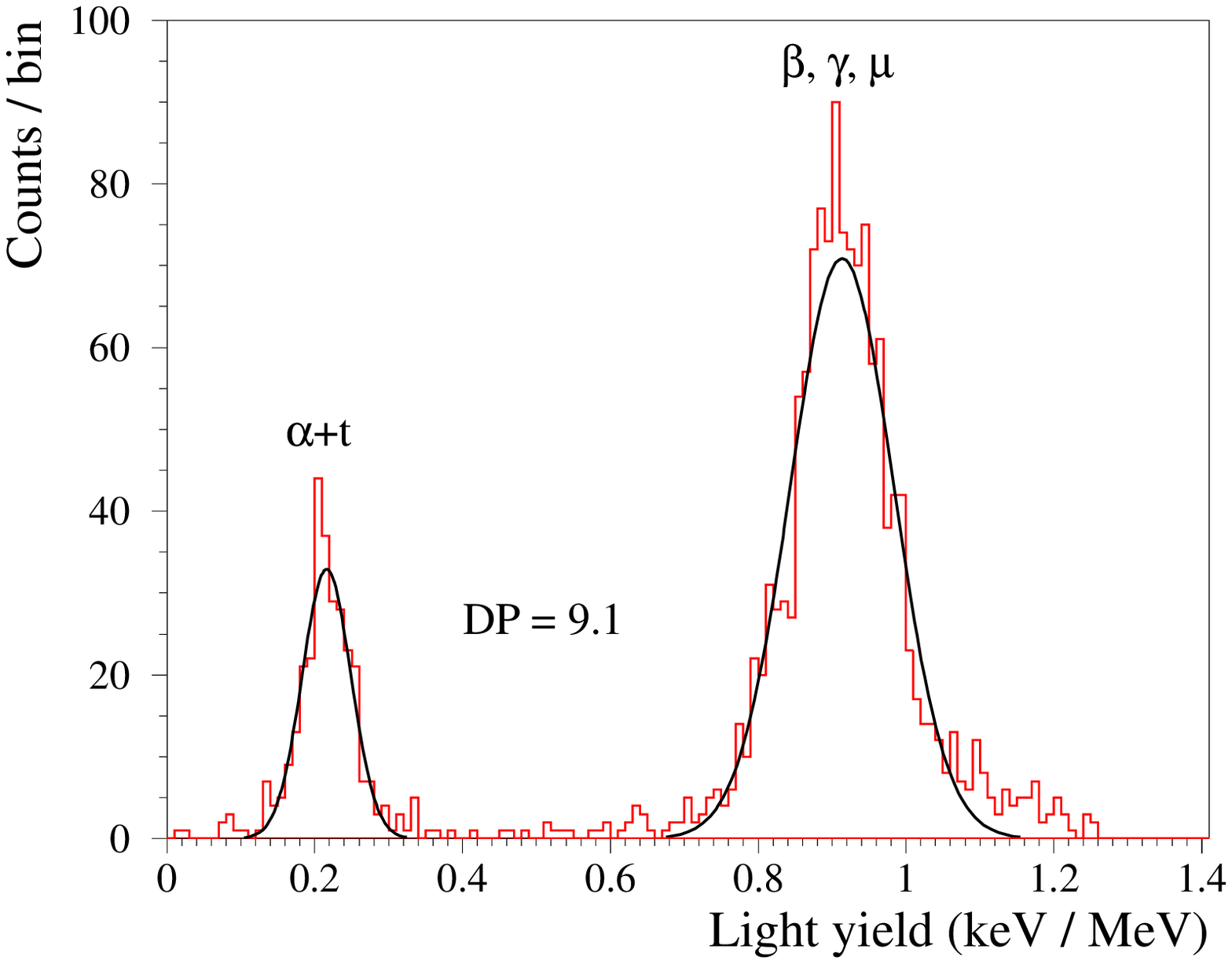}}
\end{center}
\caption{(Top) Scatter-plots of the light yield parameter versus the heat energy for the data accumulated by the LMO-large (130 h) and the LMO-small (70 h) scintillating bolometers operated above ground. The left figure is adapted from \cite{Velazquez:2017}. 
(Bottom) Projections of the light yield for events with 2.5--5.5~MeV heat energy (corresponding top figures). Two distributions are fitted by a Gaussian function. The LMO-small bolometer non-linearity (see details in the text) leads to a non-Gaussian distribution of the light yield parameter for $\gamma$($\beta$) and muon events, exhibited on the right figure as a tail of events above 1.1 keV/MeV.  
The discrimination power ($DP$) for both data is around 10.}
 \label{fig:LMO_particleID}
\end{figure*}

The coincidences between the heat and the light channels of both tested Li$_2$MoO$_4$ scintillating bolometers are depicted in Fig.~\ref{fig:LMO_particleID} (Top panel), where the dependence of a light yield parameter (a light-to-heat ratio) on the heat energy release is shown. This data representation shows $\gamma$($\beta$) and muon events clearly separated from populations with higher ionization density, sum of triton and alpha particles emitted in reactions of the neutron capture on $^6$Li and nuclear recoils as results of the scattering of the ambient neutrons on elements of the Li$_2$MoO$_4$ targets. A large amount of the muon- and neutron-induced events in the data are caused by the sea-level location of the experimental set-up (e.g., \cite{Bekker:2016,Barabash:2016}). Furthermore, the $^6$Li(n,t)$\alpha$ reaction with a $Q$-value of 4784~keV is expected response of $^6$Li-containing scintillating bolometers to neutrons because of $\sim$1~kb cross-section for thermal neutrons and $\sim$3~b of a broad resonance around 240 keV (e.g., see in \cite{Bekker:2016,Armengaud:2017}). In view of the lack of long calibration measurements with a $^{232}$Th source, we were not able to minimize the heat channel non-linearity leading to a prominent non-linearity of the LMO-small light yield for $\gamma$($\beta$)’s and muons. Moreover, the unaccounted non-linearity of the LMO-small detector is responsible for the underestimation of the heat energy of the thermal neutron capture events which are expected to give $\sim10$\% larger signal in the electron equivalent energy scale than the nominal $Q$-value of the reaction (the so-called thermal quenching; e.g., see details in \cite{Armengaud:2017}), as it is seen in the LMO-large data. The difference in the energy resolution of the Li$_2$MoO$_4$ cryogenic detectors based on the cracked LMO-large and the perfect quality LMO-small samples is evident from the comparison of the spread of the monoenergetic distributions ($\alpha$+t and/or 2615 keV $\gamma$ quanta of $^{208}$Tl) present in Fig.~\ref{fig:LMO_particleID} (Top panel). 

The efficiency of alpha background rejection was estimated by calculating the separation between the light yield distributions of $\alpha$- and $\beta$-like events registered in the 2.5--5.5~MeV energy interval, as it is illustrated in Fig.~\ref{fig:LMO_particleID} (Bottom panel). The separation is characterized by the so-called discrimination power defined as $DP = \left| \mu_{\beta} - \mu_{\alpha} \right| / \sqrt{\sigma_{\beta}^2 + \sigma_{\alpha}^2}$, where $\mu$ and $\sigma$ are the mean values and the corresponding standard deviations respectively. Taking into account the similar scintillation light yield and the light detectors performance of the tested scintillating bolometers (see Table~\ref{tab:performance}), we got similar $DP$ values around 10, which means 10$\sigma$ $\alpha$ rejection (the $DP$ meaning is illustrated by Fig.~5 in \cite{Armengaud:2017}). The ratio of the light yield of $\alpha$+t events to the one of $\gamma(\beta)$'s can be used to estimate the quenching factor ($QF$) of the scintillation signal emitted by $\alpha$'s in respect to $\beta$'s of the same energy. The results for both detectors given in Table~\ref{tab:performance} are rather similar, $\sim$0.24. It should be noted that the scintillation of $\alpha$ particles is expected to be slightly more quenched than that of $\alpha$+t events, therefore the $DP$ value for pure $\alpha$ events would be even higher than the results achieved with the $\alpha$+t distribution.

\begin{figure}[htb]
\begin{center}
\resizebox{0.49\textwidth}{!}{\includegraphics{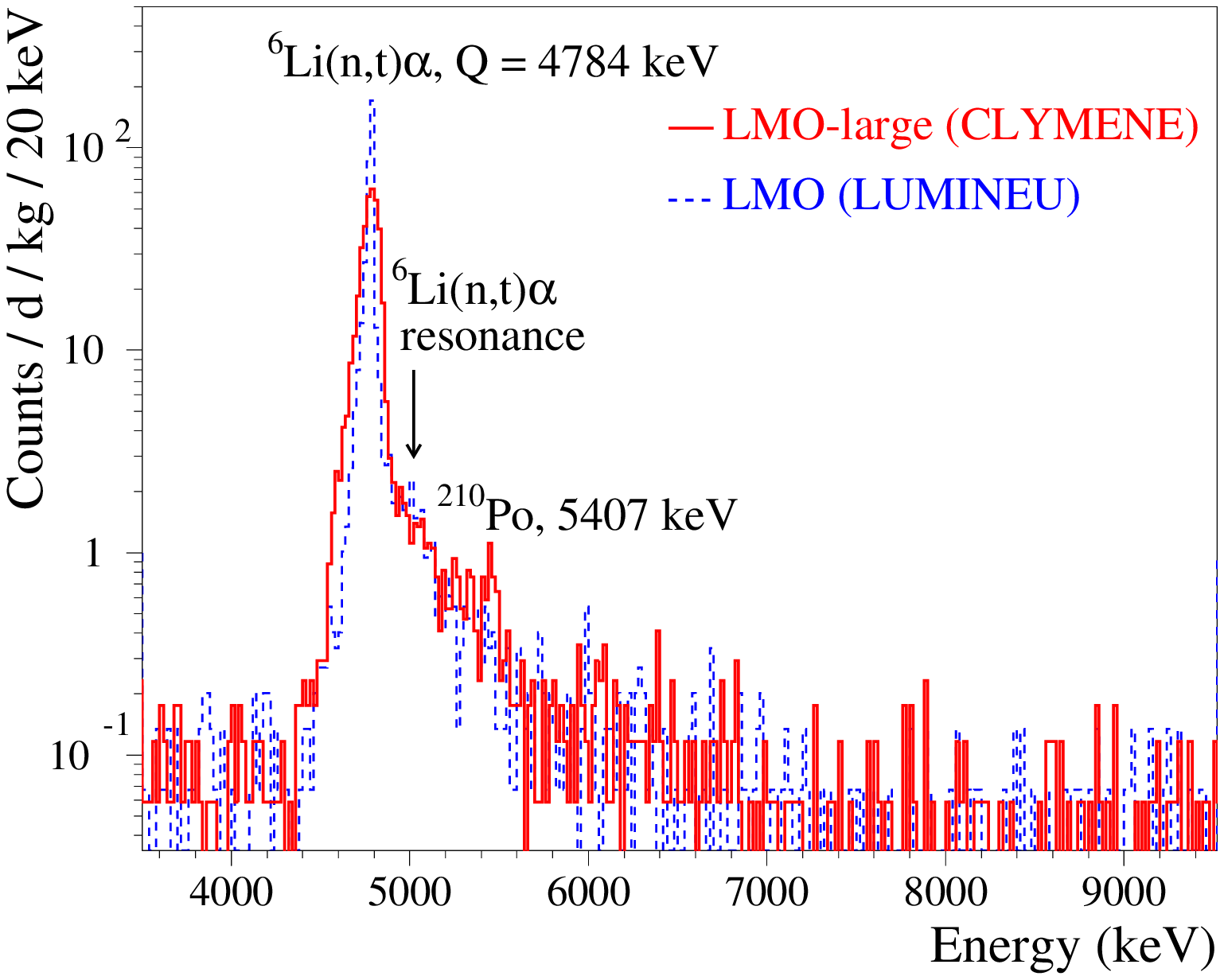}}
\resizebox{0.49\textwidth}{!}{\includegraphics{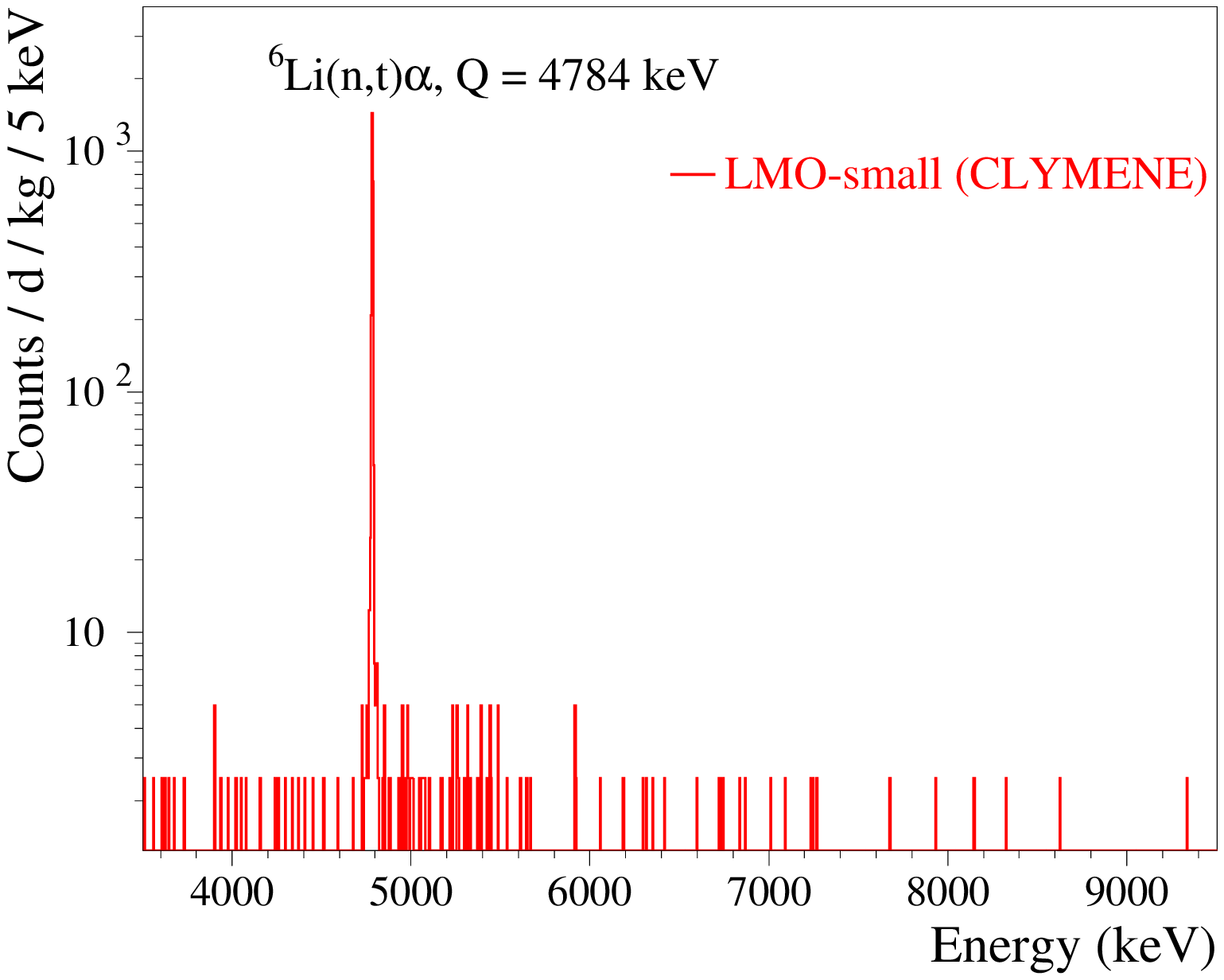}}
\end{center}
\caption{(Left) The energy spectra of $\alpha$ events accumulated by two cryogenic detectors based on similar size Li$_2$MoO$_4$ single crystals developed within the CLYMENE (LMO-large, 158 g, 130 h of data taking) and LUMINEU (151 g, 118 h, \cite{Bekker:2016}) projects. Both bolometers were operated in the same cryostat located above ground. The peculiarities associated to the internal radioactive contamination by $^{210}$Po and interactions of $^6$Li with ambient neutrons are indicated. (Right) The $\alpha$ background accumulated by the 13.5-g Li$_2$MoO$_4$ bolometer LMO-small over 144 h of measurement at sea level cryogenic set-up.}
 \label{fig:LMO_alpha}
\end{figure}

\begin{table}[htb]
\caption{Radioactive contamination of the starting materials and the first large Li$_2$MoO$_4$ scintillator  developed withing the CLYMENE project. The results for the Li$_2$MoO$_4$ crystal are extracted from the aboveground bolometric measurements with two samples produced from the same boule. The radiopurity of LUMINEU natural and $^{100}$Mo-enriched Li$_2$MoO$_4$ crystals produced by the recrystallization and operated as bolometers at underground laboratories is given for the comparison. The upper limits are given at 90\% C.L., while the errors are estimated at 68\% C.L. The $^{40}$K activity in the studied crystals is estimated from the mass-spectrometry trace element analysis of the samples cut from the top and the bottom parts of the crystal boule \cite{Velazquez:2017} (see details in the text).}
\scriptsize
\begin{center}
\begin{tabular}{|c|c|cc|cc|cc|}
 \hline
Chain & Nuclide & \multicolumn{6}{|c|}{Activity (mBq/kg)}		\\
 \cline{3-8}
~			& ~				& \multicolumn{2}{|c|}{Starting materials \cite{Velazquez:2017}} & \multicolumn{4}{|c|}{Li$_2$MoO$_4$ crystals}		\\
 \cline{3-8}
~			& ~				& MoO$_3$ 	& Li$_2$CO$_3$ & \multicolumn{2}{|c|}{CLYMENE} 	& \multicolumn{2}{|c|}{LUMINEU \cite{Armengaud:2017,Poda:2017}}	\\
~			& ~				& powder		& powder				& LMO-small			& LMO-large \cite{Velazquez:2017}	& natural & enriched  \\
\hline
$^{232}$Th	& $^{232}$Th	& $\leq12$ 	& ~ 				& $\leq0.50$	& $\leq0.18$	& $\leq0.021$ 	&	$\leq$ (0.003--0.011) 	\\
~						& $^{228}$Ra	& ~					& 37(7) 		& ~						& ~						& ~ 						& ~ 											\\
~						& $^{228}$Th	& ~ 				& 34(5) 		& $\leq0.55$	& $\leq0.25$	& $\leq0.021$ 	& $\leq$ (0.003--0.006) 	\\
\hline
$^{238}$U   & $^{238}$U		& $\leq37$ 	& ~ 				& $\leq0.72$	& $\leq0.23$  & $\leq0.048$ 	& $\leq$ (0.005--0.011)		\\
~           & $^{226}$Ra	& $\leq12$ 	& 272(14) 	& $\leq0.50$	& $\leq0.30$  & $\leq0.018$ 	& $\leq$ (0.003--0.011) 	\\
~           & $^{210}$Po	& ~ 				& ~ 				& $\leq1.7$ 	& 1.4(1)			& 0.19(4) 			& 0.020(6)--0.45(3)				\\
\hline
~    				& $^{40}$K		& 17(3) 		& 10(2)			& $\sim10$		& $\sim10$		& $\leq12$ 			& $\leq3.5$ 							\\
 \hline
\end{tabular}
\end{center}
\label{tab:radiopurity}
\end{table}
\normalsize

Thanks to highly-efficient $\alpha$/$\beta$ separation achieved by the Li$_2$MoO$_4$ scintillating bolometers, we selected the $\alpha$ events to investigate the radioactive contamination of the crystals to be exhibited by $\alpha$-active radionuclides from U/Th families. The performed analysis is similar to one described in details in \cite{Armengaud:2015}. As it is seen in Fig.~\ref{fig:LMO_alpha} (Left), the $\alpha$ background counting rate of the LMO-large detector is almost undistinguished from the data of the same-size radiopure Li$_2$MoO$_4$ crystal developed by LUMINEU and tested aboveground in the same conditions  \cite{Bekker:2016} and then underground (LMO-1 in \cite{Armengaud:2017}). The only clear difference in the spectra of Fig.~\ref{fig:LMO_alpha} (Left) is an excess of events around 5.4~MeV which indicates the $^{210}$Po content in the LMO-large sample, most probably caused by the $^{210}$Pb contamination \cite{Armengaud:2015,Armengaud:2017,Poda:2017}. The $^{210}$Po activity in the LMO-large crystal is estimated to be 1.4(1)~mBq/kg, while the activity of other $\alpha$ radioisotopes of U/Th chains is below 0.2--0.3~mBq/kg (see Table~\ref{tab:radiopurity}). In spite of the improved energy resolution of the LMO-small detector (Fig.~\ref{fig:LMO_alpha}, Right), an order of magnitude lower accumulated exposure prevents to get more stringent limits on U/Th content than ones derived from the LMO-large data (Table~\ref{tab:radiopurity}). In order to improve the detection sensitivity to the U/Th content in the CLYMENE crystals, underground bolometric measurements are needed to avoid the dominant neutron-induced background in the alpha band of Li$_2$MoO$_4$ bolometers operated aboveground. It is worth also noting a significantly higher purity of the Li$_2$MoO$_4$ crystal than that of the starting materials (see Table~\ref{tab:radiopurity}), as it is seen from at least 100 (1000) difference of the $^{228}$Th ($^{226}$Ra) activity. It illustrates the segregation of radionuclides in the Li$_2$MoO$_4$ growth process \cite{Armengaud:2017} that can be used to further improve the crystal radiopurity by the recrystallization, adopted by LUMINEU for the production of $^{100}$Mo-enriched Li$_2$MoO$_4$ scintillators \cite{Armengaud:2017}.

Taking into account the chemical affinity of lithium and potassium, Li$_2$MoO$_4$ scintillators can potentially be contaminated by $^{40}$K (see \cite{Armengaud:2017} and references therein). Moreover, as it was mentioned in \cite{Armengaud:2017}, a $^{40}$K content on the level of 60~mBq/kg would significantly contribute to the background in the region of interest of a Li$_2$MoO$_4$-based neutrinoless double-beta decay cryogenic experiment due to pile-ups of the $^{40}$K and $^{100}$Mo two-neutrino double-beta decay events. Therefore, it is rather important to control and minimize a possible $^{40}$K contamination of Li$_2$MoO$_4$ crystals. Unfortunately, we are not able to estimate the $^{40}$K activity from the present data, because the sensitivity of aboveground bolometric measurements to the $^{40}$K content is well above 0.01~Bq/kg due to the dominant environmental background\footnote{For instance, one can compare the results of aboveground \cite{Bekker:2016} and underground \cite{Armengaud:2017} low-temperature tests of a 0.15-kg Li$_2$MoO$_4$ scintillating bolometer.}. However, according to the results of a glow-discharge mass spectrometry analysis of two samples cut just below the seed and at the end of the boule \cite{Velazquez:2017}, the $^{40}$K activity in the LMO-large crystal is expected to be less than 17(3) mBq/kg but no lower than 6(1) mBq/kg (about 10 mBq/kg of $^{40}$K in the LMO-small sample due to its closer position to the seed). The 0.01~Bq/kg $^{40}$K content is reasonable acceptable and, certainly, it can be further reduced by the purification of the starting materials and/or the boule recrystallization to be comparable to the purity level of LUMINEU enriched crystals (less than few mBq/kg of $^{40}$K) \cite{Armengaud:2017}.

\section{Conclusions}

We report the encouraging results of bolometric tests of two lithium molybdate scintillators (158 and 13.5 g) produced from 230 g crystal boule, a first large-mass Li$_2$MoO$_4$ single crystal developed within the CLYMENE project aiming at realization of the new crystal production line for a tonne-scale double-beta decay cryogenic experiment CUPID. The Li$_2$MoO$_4$ cryogenic detectors demonstrate possibilities to achieve 
a high energy resolution, particularly for 2 keV FWHM baseline noise of a small bolometer, the measured energy resolution in the energy range of 0.2--5~MeV is 2--7 keV FWHM. Thanks to high scintillation properties (e.g., one of the highest light yield, 0.97~keV/MeV, ever measured with Li$_2$MoO$_4$ scintillating bolometers), the coupling of a standard performance light detector (with 0.2--0.3 keV FWHM noise) to Li$_2$MoO$_4$ bolometers allowed to get the rejection of $\alpha$-induced background on the level of 10$\sigma$. The material also exhibits potentially high radiopurity; only 1.4~mBq/kg of $^{210}$Po has been detected, while the activity of other radionuclides from U/Th chains is below 0.2--0.3~mBq/kg. Thanks to the observed segregation of $^{228}$Th and $^{226}$Ra during the Li$_2$MoO$_4$ growth, the crystal radiopurity can be further improved by the recrystallization. Both performance and radiopurity are similar to the results achieved with Li$_2$MoO$_4$ detectors recently developed by LUMINEU project. These results prove the prospects of the new R\&D on Li$_2$MoO$_4$ scintillator to develop high quality large-volume crystals needed for the realization of cryogenic double-beta detectors with high bolometric and spectroscopic performance, and very low level of radioactive contamination. The R\&D is ongoing to develop a 0.5 kg crystal boule which would be possible thanks to the comparison of the results of the numerical simulations and the Li$_2$MoO$_4$ crystal growth attempts.

\section{Acknowledgements}

This work is part of the CLYMENE project funded by the Agence Nationale de la Recherche (ANR, France; ANR-16-CE08-0018). A contribution came from the ANR-funded LUMINEU project (ANR-12-BS05-004-04) is also acknowledged. A.S.Z. is supported by the ``IDI 2015'' project funded by the IDEX Paris-Saclay (ANR-11-IDEX-0003-02).



\begin{thebibliography}{10}

\bibitem{Wang:2015a} G.~Wang, et~al., {CUPID}: {CUORE} ({C}ryogenic {U}nderground {O}bservatory for {R}are {E}vents) {U}pgrade with {P}article {ID}entification, arXiv:1504.03599.

\bibitem{Vergados:2016}
J.~D. Vergados, H.~Ejiri, F.~\v{S}imkovic, Neutrinoless double beta decay and neutrino mass, Int. J. Mod. Phys. E 25 (2016) 1630007.

\bibitem{Alduino:2017}
C.~Alduino, et~al., First results from {CUORE}: A search for lepton number
  violation via $0\nu\beta\beta$ decay of {$^{130}$Te}, Accepted for publication in Phys. Rev. Lett. (2018), 
  arXiv:1710.07988.

\bibitem{Wang:2015b}
G.~Wang, et~al., R\&{D} towards {CUPID} ({CUORE} {U}pgrade with {P}article
  {ID}entification), arXiv:1504.03612.

\bibitem{Poda:2017a}
D.~Poda and A.~Giuliani, Low background techniques in bolometers for double-beta
  decay search, Int. J. Mod. Phys. A 32 (2017) 1743012.

\bibitem{Armengaud:2017}
E.~Armengaud, et~al., Development of {$^{100}$Mo}-containing scintillating
  bolometers for a high-sensitivity neutrinoless double-beta decay search, Eur.
  Phys. J. C 77 (2017) 785.

\bibitem{Velazquez:2017}
M.~Vel\'azquez, et~al., Exploratory growth in the {Li$_2$MoO$_4$-MoO$_3$}
  system for the next crystal generation of heat-scintillation cryogenic
  bolometers, Solid State Sci. 65 (2017) 41.

\bibitem{Mancuso:2014}
M.~Mancuso, et~al., An aboveground pulse-tube-based bolometric test facility
  for the validation of the {LUMINEU} {ZnMoO$_4$} crystals, J. Low Temp. Phys.
  176 (2014) 571--577.

\bibitem{Bekker:2016}
T.~Bekker, et~al., Aboveground test of an advanced {Li$_2$MoO$_4$}
  scintillating bolometer to search for neutrinoless double beta decay of
  {$^{100}$Mo}, Astropart. Phys. 72 (2016) 38.

\bibitem{Barabash:2016}
A.~S. Barabash, et~al., First test of an enriched {$^{116}$CdWO$_4$}
  scintillating bolometer for neutrinoless double-beta-decay searches, Eur.
  Phys. J. C 76 (2016) 487.

\bibitem{Andreotti:2012}
E.~Andreotti, et~al., Production, characterization, and selection of the
  heating elements for the response stabilization of the {CUORE} bolometers,
  Nucl. Instrum. Meth. A 664 (2012) 161--170.

\bibitem{Arnaboldi:2011b}
C.~Arnaboldi, et~al., A very high performance stabilization system for large
  mass bolometer experiments, Nucl. Instrum. Meth. A 652 (2011) 306--309.

\bibitem{Gatti:1986}
E.~Gatti, P.~Manfredi, Processing the signals from solid-state detectors in
  elementary-particle physics, Riv. Nuovo Cim. 9 (1986) 1.

\bibitem{Armengaud:2015}
E.~Armengaud, et~al., Development and underground test of radiopure
  {Z}n{M}o{O}$_4$ scintillating bolometers for the {LUMINEU} $0\nu2\beta$
  project, JINST 10 (2015) P05007.

\bibitem{Poda:2017}
D.~V. Poda for LUMINEU, EDELWEISS, and CUPID-0/Mo Collaborations, {$^{100}$Mo}-enriched {Li$_2$MoO$_4$} scintillating bolometers for $0\nu 2\beta$ decay search: from {LUMINEU} to {CUPID-0/Mo} projects, 
AIP Conf. Proc. 1894 (2017) 020017.

\end{thebibliography}
\end{document}